\documentclass[a4paper,10pt]{article}

\usepackage{amsmath}
\usepackage{vmargin}
\usepackage{graphicx}
\usepackage{hyperref}
\hypersetup{colorlinks,pdftitle={H. B. Reitlinger and the origins of the
Efficiency at Maximum Power formula for Heat Engines},pdfauthor={A.~Vaudrey,
F.~Lanzetta \&
M.~Feidt},citecolor=blue,filecolor=blue,linkcolor=blue,urlcolor=magenta}

\title{\bf H. B. Reitlinger and the origins of the Efficiency at Maximum Power
formula for Heat Engines}
\author{
   	A.~Vaudrey\thanks{Corresponding author :
	\href{mailto:alexandre.vaudrey@ecam.fr}{alexandre.vaudrey@ecam.fr}}\textsuperscript{
,}\thanks{University of Lyon, ECAM, Energy Laboratory, 40 montée Saint Barthélemy, 69321 Lyon, France.}
	\and
	F.~Lanzetta\thanks{University of Franche-Comte, FEMTO-ST, UMR CNRS 6174,
	Parc technologique, 2 av. Jean Moulin, 90000 Belfort, France.}
	\and
   	M.~Feidt\thanks{University of Lorraine, Laboratory of Energetics \&
	Theoretical \& Applied Mechanics, 2 av. de la For\^et de Haye, 54518
	Vand\oe uvre-l\`es-Nancy, France.}}

\begin{document}
\maketitle

\begin{abstract}
	Even if not so ancient, the history of the heat engine efficiency at
	maximum power expression have been yet turbulent. More than a decade
	after the publication of the seminal article by \textsc{Curzon} and
	\textsc{Ahlborn} in 1975, one's rediscovered two older works by
	\textsc{Chambadal} and \textsc{Novikov}, both dating from 1957. Then,
	some years ago, the name of \textsc{Yvon} arose from a textual reference
	to this famous relation in a conference article published in 1955.
	Thanks to an historical study of French written books not anymore
	published for a long time, and since never translated in other languages,
	we bring to light in this paper that this relation was actually firstly
	proposed by \textsc{Henri B. Reitlinger} in 1929.
\end{abstract}

\section{Introduction} The heat engine efficiency at maximum power formula is
probably the most famous and important one of modern technical thermodynamics.
It pushed the latter out of the fetters of strictly equilibrium configurations,
and then closer to behaviors of actual systems, and gave birth to what is called
now the \textit{Finite Time Thermodynamics} (FTT),
\cite{Berry-2008-01,ACIE-050-2690}.  However, besides its own physical meaning,
its recent but lively history could say a lot about the sometimes sinuous path
of science. 

In this paper, thanks to the study of not anymore published for a long time
French written books, we will come back to the root of this formula. We will
follow its interesting story, up to its first publication in 1929, by
\textsc{Henri B. Reitlinger}. The goal is here to render unto Caesar that which
is Caesar's in giving to \textsc{Henri Reitlinger} the place he deserves in the
history of thermodynamics.

\section{A short history of the efficiency at maximum power formula}

\paragraph{Curzon and Ahlborn, 1975} 

Most of the references dealing with the thermal efficiency at maximum power of a
reversible heat engine operating between cold and hot heat reservoirs
respectively at temperatures $T_{\text{cold}}$ and $T_{\text{hot}}$, i.e. :
\begin{equation}
	\eta_{\text{opt}} = 1 - \sqrt{\frac{T_{\text{cold}}}{T_{\text{hot}}}}
	\label{eff-max-1}
\end{equation} cites the seminal article of \textsc{Curzon} and
\textsc{Ahlborn}~\cite{AJPh-043-0022-0024}, published in 1975. 

This article presented a theoretical analysis of the fundamental Carnot cycle,
but with each of its two isothermal heat exchange processes occurring in finite
time, \cite{AJPh-053-0570-0573}. One of the most impressive aspects of relation
\eqref{eff-max-1} is its complete independence on the specific characteristics
of concerned engine, exactly as for the well known Carnot's formula. This
fascinating result was the starting point of the \textit{Finite Time
Thermodynamics} \cite{Berry-2008-01}. Since then and during more than two
decades, $\eta_{\text{opt}}$ of relation (\ref{eff-max-1}) was fairly called
\textsc{Curzon-Ahlborn} efficiency, often noted~$\eta_{\rm CA}$.

\paragraph{Chambadal and Novikov, 1957} 

In at least two articles, \textsc{Bejan}
\cite{AJPh-062-0011,JAPh-079-1191-1218}, previously warned by one of us
\cite{Feidt-1987}, reminded to the scientific community that relation
(\ref{eff-max-1}) was actually published almost two decades before the
\textsc{Curzon} and \textsc{Ahlborn} paper, separately by two different
authors~:~\textsc{Chambadal}~\cite{Chambadal-tr-eng} and
\textsc{Novikov}~\cite{Novikov}. The \textsc{Novikov}'s article was actually
published in Russian in 1957, the same year as the \textsc{Chambadal}'s book,
but translated in English one year later. Both studies were devoted to nuclear
energy, the first one considering a Carnot reversible engine fed in heat by a
finite capacitance stream and the second one more focusing on the thermodynamic
cycle itself, see \cite{JAPh-079-1191-1218}. Even if some authors
\cite{PhRE-085-041144} have recently expressed doubts about the
\textsc{Chambadal}'s paternity on relation \eqref{eff-max-1}, mainly because of
a different model than the \textsc{Novikov} one, we insist on the identity of
both results. The aim of both studies is the same : to maximize the useful
effect of heat engines, whether this effect is expressed as work or power.

\smallskip

\textsc{Bejan} then proposed to call the concerned efficiency (\ref{eff-max-1})
``\textsc{Chambadal-Novikov-Curzon-Ahlborn} efficiency'', and to note it
$\eta_{\text{CNCA}}$.  Surprisingly, many authors still neglect this important
historical update and often cite in recent papers only
reference~\cite{AJPh-043-0022-0024}.

\paragraph{Yvon, 1955} 

More recently, \textsc{Moreau} \emph{et al}~\cite{EPhJD-062-0067}, working on
the specific situation of maximum power production by thermal engines, cited a
conference article published by \textsc{Jacques Yvon}~\cite{Yvon-1955} in 1955,
two years before the \textsc{Chambadal} and \textsc{Novikov} publications, and
containing the following strange because of harmless quotation : \begin{quote}
	``\emph{Without specifying the thermal power at which the reactor must
	operate, the mean fluid temperature can be chosen so that the usable
	power be as large as possible. The result is that it must be the
	geometric mean of the [high and low] temperatures [...]}''
\end{quote} so exactly the approach leading to equation (\ref{eff-max-1}). 

\smallskip

Even if the concerned conference was probably not dedicated to thermodynamics or
heat engines, but rather to nuclear reactor science, such casual statement is
disturbing because seemingly obvious for the author, as if it was a widely known
and accepted result. Unfortunately, publication \cite{Yvon-1955} made no
reference to previous work on this topic and leads to a dead-end for historical
research.

\smallskip

The only way remaining to investigate on the story of formula (\ref{eff-max-1})
seemed to be contained in one of the books about thermodynamics published by
\textsc{Paul Chambadal}. Sadly, none of these were translated in an other
language than French, which could partly explain the lack of spreading of results
we will discuss about in the rest. 

\begin{figure}[t!]
	\centering
	\includegraphics[width=0.7\textwidth]{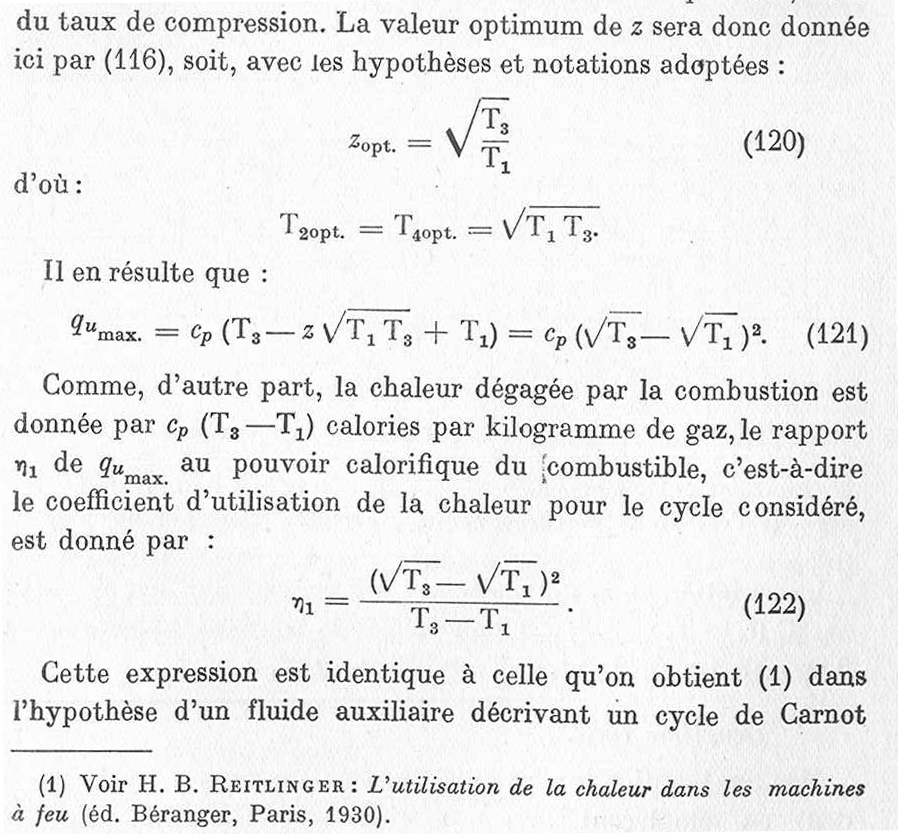}
	\caption{Extract from \textsc{Chambadal}'s book~\cite[page
	225]{Chambadal-1949-tr-eng} published in 1949 and yet containing some
	equations nowadays related to Finite Time Thermodynamics. One could also
	notice the footnote with reference to the edition of the
	\textsc{Reitlinger}'s book published in France in 1930, so one year
	later the Belgian first one.}
	\label{page-Chambadal-1}
\end{figure}

\paragraph{Chambadal, 1949} Indeed, in a book published in 1949 and dedicated to
thermodynamics of gas turbines, \textsc{Chambadal}~\cite{Chambadal-1949-tr-eng}
yet studied configurations of maximum produced work by such system, and obtained
relation (\ref{eff-max-1}) and others resulting from it, even if in a slightly
different way. On figure \ref{page-Chambadal-1} is presented a picture taken
from the page 225 of the \textsc{Chambadal}'s book \cite{Chambadal-1949-tr-eng}
from 1949, containing among others the famous relation between optimal
temperature $T_{2,\text{opt}}$ and corresponding cold and hot ones $T_1$ and
$T_3$. This relation naturally leads the author to obtain efficiency
\eqref{eff-max-1} in following pages. Fortunately, \textsc{Chambadal} cites as
reference (on the footnote of figure \ref{page-Chambadal-1}) a previous book on
this topic, published by \textsc{Henri B. Reitlinger} in 1930.
 \begin{figure}[t!]
	\centering
	\includegraphics[width=0.5\textwidth]{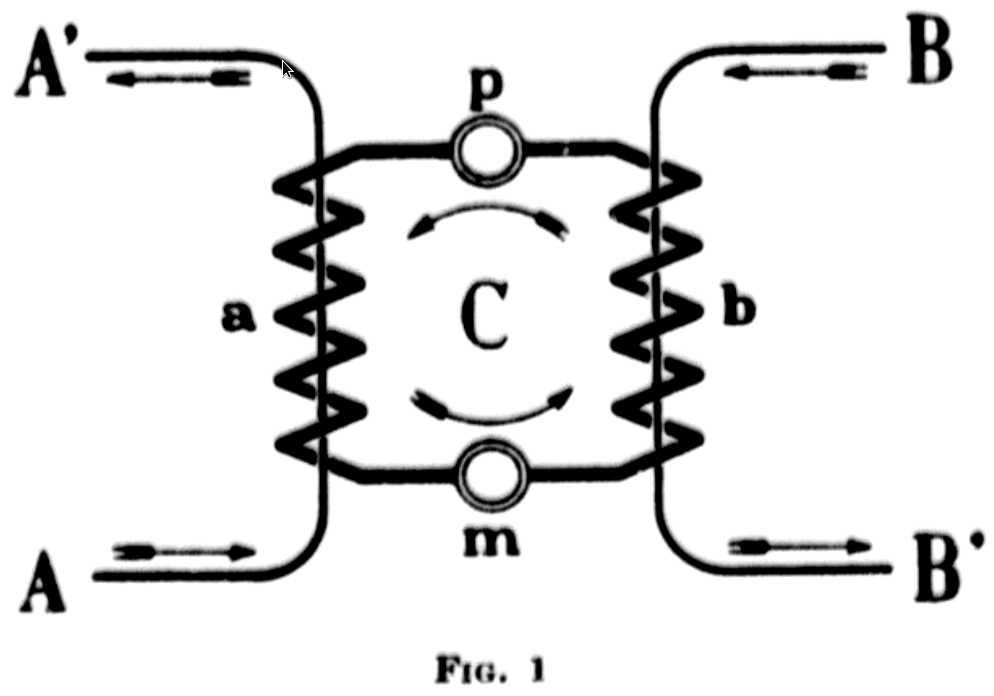}
	\caption{Representative figure of a heat engine considered as an open
	system, hot stream entering in A and exiting in A', the B-B' circuit
	being dedicated to cold stream. The heat to work conversion device is
	here indicated by letter m (for machine) and represented by circuit
	noted C. $a$ and $b$ are respectively the hot and cold heat
	exchangers and p the pump of the primary circuit \cite[page
	12]{Reitlinger-1929-ang}.}
	\label{fig-Reitlinger-1}
\end{figure}

\section{The Reitlinger book}

The concerned book, firstly published by \textsc{Henri B.
Reitlinger}~\cite{Reitlinger-1929-ang} in 1929 was dedicated to practical
designs and uses of heat engines, at this time often driven by the combustion of
coal. Such system was considered by the author as open, and composed by a heat
to work conversion device exchanging heat with hot and cold fluids thanks to
heat exchangers, as presented on figure~\ref{fig-Reitlinger-1}.  Heat exchangers
A-A' and B-B' are respectively used to feed the work conversion device C with a
hot stream produced by a combustion process, and expel the remaining cold heat
to surroundings thanks to e.g. a liquid water cooling system.

\begin{figure}[t!]
	\centering
	\includegraphics[width=0.7\textwidth]{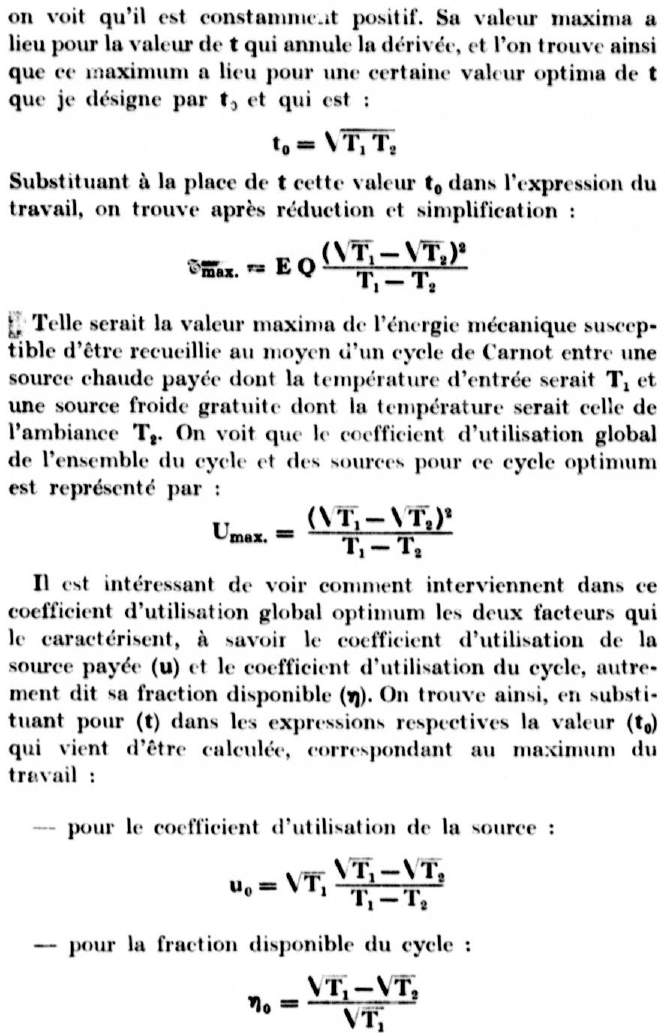}
	\caption{Page of the \textsc{Reitlinger}'s book~\cite[page
	28]{Reitlinger-1929-ang} presenting yet in 1929 the expressions of
	temperatures, produced work and relating efficiency of a heat engine
	operating at maximum power configuration. The last equation at the
	bottom of the page is actually the first one in history about the
	efficiency at maximum power.} 
	\label{page-Reitlinger-1}
\end{figure}

Describing heat engines operating principle in this way naturally leads to
consider the need of temperature differences within each hot and cold streams,
in order to exchange heat with C, the heat-to-work conversion part of the
system. \textsc{Reitlinger} logically studied the impact of such temperature
differences on the performance of the whole system, and obtained in first the
relation (\ref{eff-max-1}), as presented on the bottom of figure
\ref{page-Reitlinger-1}.  Even non French readers could clearly notice the expressions of
optimal temperature $t_0 = \sqrt{T_1 \cdot T_2}$ of the exhaust hot stream,
defined as previously explained as the geometric mean of hot and cold entrance
temperatures of stream, respectively noted $T_1$ and $T_2$. On figure
\ref{fig-Reitlinger-1}, hot stream composed by coal combustion exhaust gas
enters in A at temperature $T_1$ and exits in A' at temperature~$t$. 

In the case presented on figure \ref{page-Reitlinger-1}, cold heat exchanger
between points B and B' is supposed isothermal at temperature $T_2$, thanks to
an almost infinite thermal flow $\dot{m} \cdot c_p$ of liquid water, regarding
to the hot one composed by combustion exhaust gas flow. 

We can notice that \textsc{Chambadal}~\cite{Chambadal-tr-eng} applied exactly
the same calculation on a steam turbine in his 1957's book about nuclear power
plants. 

Replacing hot exit temperature $t$ by its optimal value $t_0$,
\textsc{Reitlinger} obtained the second equation on picture
\ref{page-Reitlinger-1}, related to the maximum amount of work such engine could
produce. Finally, when interested by the thermal efficiency expression, he
obtained for the first time the formula (\ref{eff-max-1}), written as :
\begin{equation}
	\eta_0 = \frac{\sqrt{T_1} - \sqrt{T_2}}{\sqrt{T_1}} 
\end{equation} as presented on the bottom of figure~\ref{page-Reitlinger-1}.

\section{Conclusion} History of science is well known to not be a long and quiet
river. Historical studies sometimes remind to the scientific community the
names of scientists that firstly proposed theories, which became later famous
without the name of their original creators. From time to time, a concept has
been discovered and developed by several scientists at the same time and just
few of them have let their name on the resulting theory. The ``\textsc{Betz}
limit'' of produced power by propellers or wind turbine is a typical example of
that~\cite{WE-010-0289}. 

\smallskip

More rarely, the concerned concept or theory have been published a long time
before, within a different scientific domain or in another language, and had
never been translated since. This is probably what happened for the heat engine
efficiency at maximum power, which was actually published almost half a century
before the most often cited paper from \textsc{Curzon} and \textsc{Ahlborn}.
This fact obviously doesn't withdraw any virtue from these authors and doesn't
cast any doubt on their fundamental contribution to the creation of Finite Time
Thermodynamics. With the historical perspective, the result they obtained
appears as a brilliant completion of a thought process started a long time
before, thanks to more specific and practical purpose.  Finally, it emphasize
the fundamental importance of scientific information spreading, of
collaborations between scientists reading different languages and of historical
studies once in a while.

\section*{Acknowledgments} Authors want to thank the Biblioth\`eque nationale de
France (BnF) for the digitizing of the \textsc{Reitlinger}'s book, that allowed
the present study and paper.

\bibliographystyle{unsrt}
\bibliography{biblio2}

\end{document}